\documentclass[aps, preprintnumbers,amsmath,amssymb,
eqsecnum, twocolumn, tightenlines]{revtex4}

\usepackage{dcolumn}
\usepackage{bm}
\usepackage{graphicx}
\usepackage{multirow}
\usepackage{color}
\usepackage{graphicx}
\usepackage[utf8]{inputenc}
\usepackage{pbox}
\usepackage{graphicx, fancybox}
\usepackage{amsmath,amssymb}
\usepackage[colorlinks=true, pdfstartview=FitV, linkcolor=red,
citecolor=blue, urlcolor=blue]{hyperref}
\usepackage{soul}
\usepackage{url}
\usepackage[normalem]{ulem}

\keywords{electromagnetic field at RHIC; elliptic flow; QGP}

\begin{document}
	\title{Electromagnetic field evolution in relativistic heavy ion collision\\ and its effect on flow of particles}
	
	\author { Tewodros Gezhagn$^{1,2}$ and A. K. Chaubey$^1$ }
	\affiliation { $^1$ Department of Physics, Addis Ababa University, P.O. Box 1176, Addis Ababa, Ethiopia}
	\affiliation { $^2$ Department of Physics, Aksum University, P.O. Box 7080, Aksum, Ethiopia}
	\date{\today}
	\begin{abstract}
		
	We compute the electromagnetic fields generated in relativistic heavy-ion collisions using the iEBE-VISHNU framework. We calculated the incremental drift velocity from the possible four sources of the electric force (coulomb, Lorentz, Faraday and Plasma based) on the particles created. The effect of this external electromagnetic field on flow harmonics of particles was investigated, and we found out that the flow harmonics values get suppressed and rouse in non uniform fashion through out the evolution. More precisely, a maximum of close to three percent increase in elliptic flow was observed. We also found mass more dominant factor than charges for the change in flow harmonics due to the created electromagnetic field. On the top of that the magnetic field perpendicular to the reaction plane is found to be sizable while the different radial electric forces were found to cancel out each other. Finally, we found out that the inclusion of electromagnetic field affects the flow of particles by suppressing or rising it in non uniform fashion through out the evolution.
			
	\end{abstract}
	\maketitle
	\section{Introduction}

	Understanding how matter behaved at the beginning of the universe, 
	by creating and studying the Quark Gluon Plasma is the primary purpose of 
	the relativistic heavy-ion collisions experiment.~\cite{Nasim} From the many possible 
	signatures of the quark-gluon plasma, the very convincing evidences which led to the 2005's announcement, the QGP had been  discovered at RHIC ~\cite{2008jna,PefectLiquid}, came from the combination of three observations: the measurements of strong anisotropic collective flow, valence quark number scaling of the elliptic flow $v_2$ , and jet quenching.~\cite{2008jna,Jacobs:2007dw} Elliptic flow is a measure of how the energy$,$ momentum and number of created particles are not uniform with direction. The Fireballs produced by the little Bangs at RHIC and the LHC undergo explosive collective expansion cooling down rapidly and finally fragments into thousands of free-streaming hadrons~\cite{Heinz:2013wva}. Theoretically, relativistic hydrodynamics has established itself as an indispensable component in modeling the collective dynamics of the Quark-Gluon Plasma (QGP) produced in relativistic heavy ion collisions ~\cite{Song:2008UH, Song:2011BH,Luzum:2011, QGP-Lee:1974ma,QGP-Collins:1974ky}. The first and principal observable that has been extensively used for extracting information about the QGP from heavy ion 
	collisions is the collective flow of various particles, specially the elliptic flow~\cite{song:2007cvh}. This implies that any elliptic flow related studies have the chance to clear the road to understand QGP well; and answer to what affects elliptic flow is crucial.
	
	
	\begin{figure}[h!]
		\vspace{0.2in}
		\begin{center}
		   \includegraphics[scale=0.3]{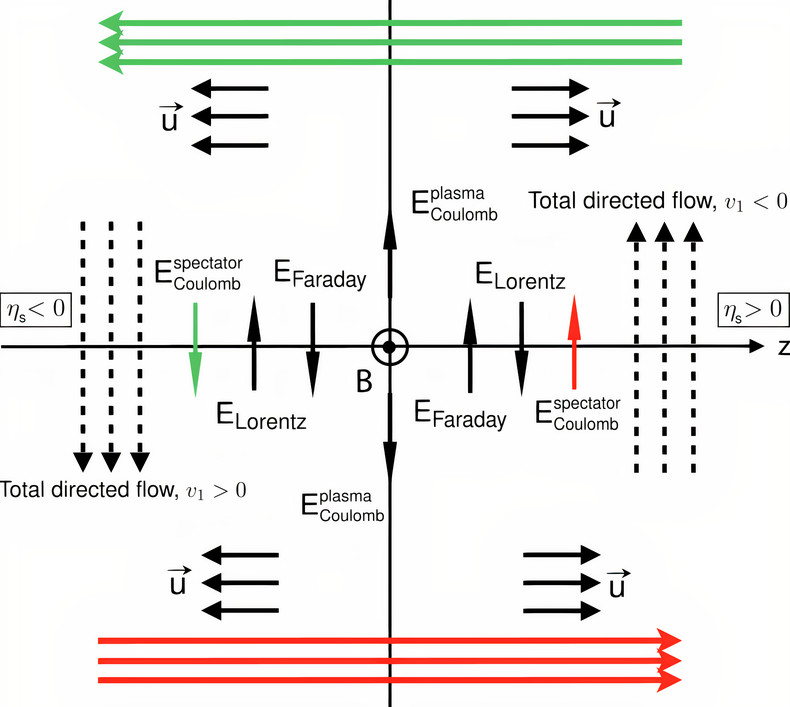}
		\end{center}
		  \caption{(Color online) Schematic illustration of how the electromagnetic fields in a heavy ion collision result in a directed flow of electric charge ~\cite{Shen:2018GKM}}
		\label{figschema}
	\end{figure}
	It is obvious that there are various parameters affecting flow harmonics such as starting time, initial energy profile, initial flow velocity with the viscous stress tensor, equation of state (EOS), specific shear viscosity, kinetic decoupling temperature and many more ~\cite{Shen:2014lye}. However, in this research, we picked the electromagnetic field whose effect is not widely considered by many of the hydro models. In order to address this problem, our model set-up followed three basic steps. The first step is simulating the dynamical evolution of the medium produced in $Pb-Pb$ collision at $ecm =$ ${2.76 TeV}$ using the iEBE-VISHNU frame-work. The second step is computing the drift velocity from the electromagnetic field applying the Maxwell's equation using the information we obtained from the hydrodynamic simulator. The third step is finding out the flow harmonics by injecting the drift velocity which is deduced in the framework. As a result, the first thing that comes is exploring the sources of the electric fields, and it has been stated as follows.
		
	\section{The Elecro-magnetic field}
	
	The possible origins of the created electric force are four. The first is from the positively charged spectators that fly away the collision zone. There is an electric force on the charged Plasma produced there. This is a direct coulomb source field. Secondly, from the spectator nucleons and the charge density deposited in the plasma. We have a non-vanishing outward-pointing component of the electric field already in the lab frame. 
	
	Thirdly, there is Lorentz force that can contribute to the directional pushes particles experience. This is because of the perpendicular magnetic field $\vec B$ created in the system. Finally, there is an induced electric current because of the change in magnetic field, {\tt Faraday}. A more detailed explanation is found in~\cite{Shen:2018GKM}. These four sources of electric forces : the {\tt coulomb} $(\vec E_C)$, {\tt Lorentz:} ($\vec E_L$), {\tt Faraday:} $(\vec E_F)$ and  {\tt Plasma based}, are illustrated in Fig.~\ref{figschema}. Inorder one compute the incremental drift velocity $\vec{v}_\mathrm{drift}$ caused by the electromagnetic forces, the electric $(\vec E)$  and magnetic fields $(\vec B)$ evolved should be known from the governing Maxwell's equation.  
	
	
	\section{Methods}
	
	As we all know, there is no unique theoretical tool to describe the whole heavy ion collision process from the very beginning till the end and this forces as set up a model. We have used a similar model-set up as in~\cite{Shen:2018GKM} except we extend the drift velocity for more charge types and evaluate the effects for different particles. Following the set-up, we simulate the dynamical evolution of the medium produced in $Pb-Pb$ collision at $ecm =$ ${2.76 TeV}$ using the iEBE-VISHNU frame-work and the electromagnetic fields evolution which is computed alone. In general, according to the model set up, there are three basic jobs to be done. The first is simulating the dynamical evolution using same Monte-Carlo–Glauber initialization. Second one is compute the additional velocity induced by the electromagnetic field alone and add this with velocity from the dynamical system itself. In the last step we injected this final velocity to the particle sampler in the framework. \\

		\begin{figure*}
		\centering
		\includegraphics[scale=0.45]{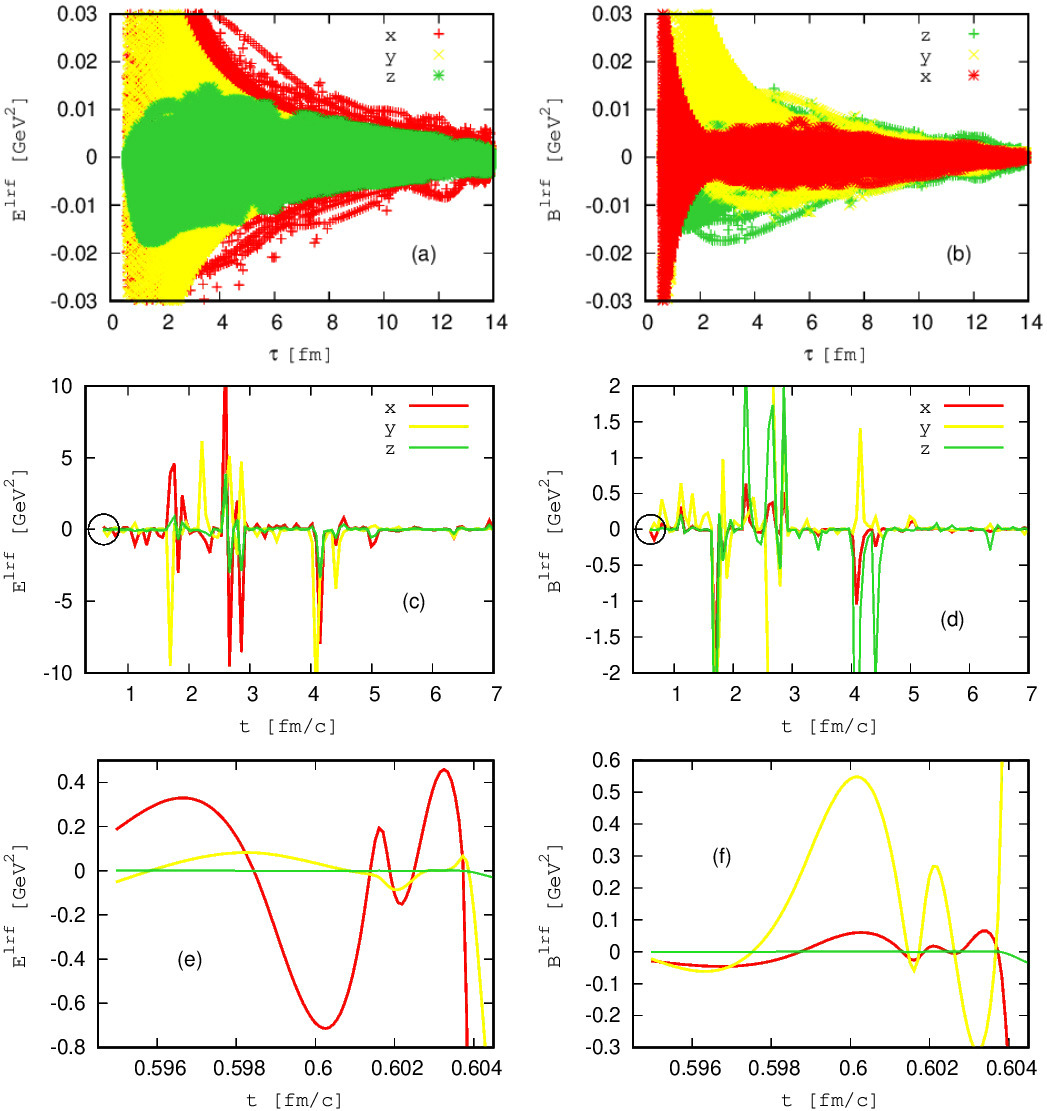}	
		\caption{(Color online) (a) : Illustrates evolution of the three components of the electric fields in the local fluid rest frame at points on the freeze-out surface created in relativistic Pb+Pb collision with collision energy of $\sqrt{s}$ = 2.76 ATeV. The calculations were made at centrality ranging from 20-30 \% (impact parameters in the range 6.24 fm $<$ b $<$ 9.05 fm) and $ \eta_s$ ranging from -7.9 to 7.9 where z is the reaction axis while x and y represents the transverse and longitudinal directions respectively. (b) : Illustrates evolution of the three components of the magnetic fields in the local fluid rest frame at points on the freeze-out surface created in relativistic Pb+Pb collision with collision energy of $\sqrt{s}$ = 2.76 ATeV. The calculations were made at centrality ranging from 20-30 \% (impact parameters in the range 6.24 fm $<$ b $<$ 9.05 fm) and $ \eta_s$ ranging from -7.9 to 7.9. (c): The three components of the electric fields versus the local time in the fluid rest frame at a fixed space time rapidity of $ \eta_s$ = 0. (d) : Shows the three components of the magnetic fields versus the local time in the fluid rest frame at a fixed space time rapidity of $ \eta_s$ = 0. (e) : Illustrates the magnified version of the circled part from figure (b), and shows the three components of the electric fields evolution during the very early propagation time in the fluid rest frame at a fixed space time rapidity of $ \eta_s$ = 0. (f) : Illustrates the magnified version of the circled part from figure (d), and shows the three components of the magnetic fields evolution during the very early propagation time in the fluid rest frame at a fixed space time rapidity of $ \eta_s$ = 0.}
					
		\label{EMevolution.codes}
	\end{figure*}
	
	\subsection{The iEBE-VISHNU frame-work}
	
	In this hybrid package, there are specific codes simulating each stage of the evolution and a script to link all the individual programs together. The four major components of the package are :
	
	\begin{enumerate}
		\item {\tt superMC }: ( the initial condition generator )
		This code generates fluctuating initial conditions according to Monte-Carlo Glauber and Monte-Carlo Kharzeev-Levin-Nardi (KLN)  $"$gluon saturation$"$ models. In this work, we have used the Monte-Carlo Glauber model initialization which assumes that the initial energy density in the transverse plane is proportional to the wounded nucleon density.
		
		\item {\tt VISHNew }: ( viscous hydrodynamic simulator )
		Viscous Israel Stewart hydrodynamics is a (2+1)-d viscous hydrodynamic simulation for relativistic heavy-ion collisions.
		It solves the equation of motion for second order viscous hydrodynamics Israel-Stewart equations with a given equation of state (EOS). VISHNew supports three versions (s95p-v0-PCE, s95p-v1, and s95p-v1-PCE) of the lattice-based equation of states by applying different implementations of partial chemical equilibrium in the hadronic phase. We have used s95p-v1-PCE for determining the pressure in the fluid, by first solving for the local energy density and velocity of the fluid cell.	
		
		\item {\tt iSS }: ( a particle sampler )
		iSS is an “event generator” which generates a complete collision event of emitted hadrons from Cooper-Frye freeze-out procedure similar to the events created in the experiment.
		
		\item {\tt UrQMD }: ( afterburner ) 
		This is a hadron cascade simulator ideally suited for the description of the dynamics of a system of hadrons, both in and out of equilibrium.

	\end{enumerate}
\begin{figure*}
	\centering
	\includegraphics[scale=0.6]{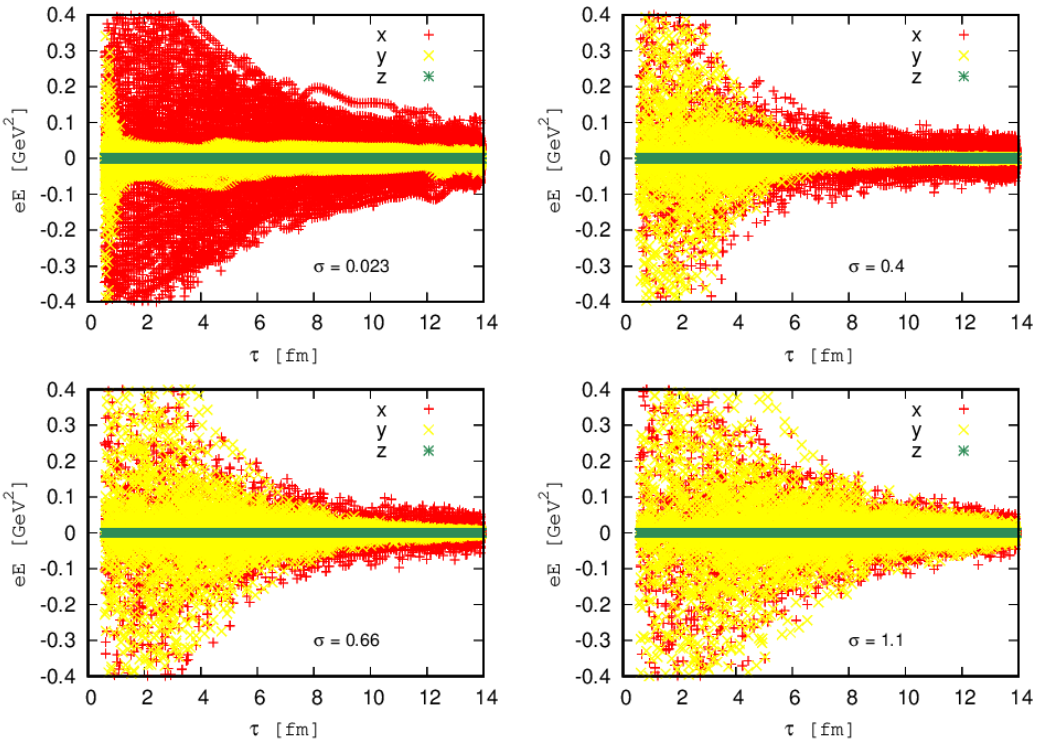}
	
	\caption{(Color online) The electric field evolution created in a system of relativistic heavy-ion collision, after a Pb+Pb collision with 20-30 \% centrality at collision energy $\sqrt s$ = 2.76 ATeV on the choice of different electrical conductivity for the Maxwell equations. }
	\label{conductivity}
\end{figure*}

	\section{Solving Maxwell's Equation}
	
	The electromagnetic fields generated by a point-like charge moving in the +z-direction with velocity $v$ is governed by the following wave equations~\cite{Shen:2018GKM}.  
	\begin{eqnarray}
	\nabla^2\vec{B} - \partial_t^2 \vec{B} - \sigma  \partial_t \vec{B} &=& - \vec{\nabla} \times \vec{J}_\mathrm{ext} \\    \nabla^2\vec{E} -\sigma^2_t \vec{E} - \sigma \partial_t \vec{E} &=& \frac{1}{\epsilon} \vec{\nabla} \rho_\mathrm{ext} + \partial_t \vec{J}_\mathrm{ext}\  \label{eq8}
	\end{eqnarray}
	
	The external charge and current sources for the electromagnetic fields~\cite{Shen:2018GKM} are functions of rapidity and given by :
	
	\begin{equation}
	\rho_\mathrm{ext} (\vec{x}_\perp, \eta_s) = \rho^+_\mathrm{ext} (\vec{x}_\perp, \eta_s) + \rho^-_\mathrm{ext} (\vec{x}_\perp, \eta_s)
	\end{equation}
	\begin{equation}
	\vec{J}_\mathrm{ext} (\vec{x}_\perp, \eta_s) = \vec{J}^+_\mathrm{ext} (\vec{x}_\perp, \eta_s) + \vec{J}^-_\mathrm{ext} (\vec{x}_\perp, \eta_s)
	\end{equation}
	
	Solving the above wave equations is easy since we have considered the electrical conductivity of the QGP to be constant, but making it time varying, which in reality is, because of its dependence on the temperature of the system would make it very hard and is what we anticipate to work on next. So we took four different $\sigma$ values $\sigma = 0.023$ fm$^{-1}$, $\sigma = 0.4$ fm$^{-1}$, $\sigma = 0.66$ fm$^{-1}$ and $\sigma = 1.1$ fm$^{-1}$ to have a chance of analyzing its effect. And the numerical code for calculating the electromagnetic fields is the same as in~\cite{Shen:2018GKM} which is found at \url{https://github.com/chunshen1987/Heavy-ion_EM_fields}.  For the evolution of the relativistic viscous hydro-dynamics, we chose the s95p-v1-PCE equation of state from Ref.~\cite{Petreczky:2010}. After we knew the field evolution, we calculated the drift velocity.
	
	\begin{table*}[ht]
		\centering
		\begin{tabular}{*{16}{|c}|}
			\hline
			\hline
			No. & Particle type          & Pt $(GeV)$ & $V_2$ Theory & $V_2$ with EM & $\Delta V_2$ &  \pbox{30cm}{Percentage Increase \\ of $ V_2$ (in \% ) due to \\ the electromagnetic field} \\	
			\hline
			\hline
			1   & \multirow{14.5}{*}{ Proton} &  $0.00717752$           & -$1.79444*10^-$$^6$   & -$1.81221*10^-$$^6$         & -$1.777*10^-$$^8$                        & 0.990285         \\
			2   &                         & $0.0935073$          & -$2.93797*10^-$$^4$  &  -$290903*10^-$$^4$          & $2.89379*10^-$$^6$                       & -0.984963           \\
			3   &                         & $0.0378994$           & -$5.04515*10^-$$^5$    & -$4.99594*10^-$$^5$          & $4.92179*10^-$$^7$                        & -0.975547          \\ 
			4   &                            & $0.282125$            & -$1.61245*10^-$$^3$   & -$1.63177*10^-$$^3$        & -$1.93163*10^-$$^5$                       & 1.19795           \\
			5   &                         & $0.174611$            & -$8.91892*10^-$$^4$   & -$8.82765*10^-$$^4$         & $9.12714*10^-$$^6$                          & -1.02334          \\
			6   &                         & $0.581994$            & $2.54867*10^-$$^3$    & $2.51128*10^-$$^2$           & -$3.73926*10^-$$^5$                      & -1.46714          \\
			7   &                         & 0.417334            & -$1.26594*10^-$$^3$    & -$1.2359*10^-$$^3$            &   $3.00437*10^-$$^5$                       & -2.37322         \\
			
			8   &                         & $0.778479$          & $1.12569*10^-$$^2$  &  $1.12195*10^-$$^2$          & -$3.74261*10^-$$^5$                       & -0.332473            \\
			9   &                         & $1.01002$           & $2.361811*10^-$$^2$    & $2.35998*10^-$$^2$          & -$1.82688*10^-$$^5$                        & -0.077351           \\ 
			10   &                         & $1.28111$            & $3.65891*10^-$$^2$   & $3.65551*10^-$$^2$         & -$3.39672*10^-$$^5$                          & -0.0928342           \\
			11   &                         & $1.59819$            & $4.76291*10^-$$^2$   & $4.75076*10^-$$^2$         & -$1.21351*10^-$$^4$                          & -0.254785         \\
			12   &                         & $1.97107$            & $5.54413*10^-$$^2$    & $5.52049*10^-$$^2$           & -$2.36478*10^-$$^4$                      & -0.426537          \\
			13   &                         & 2.41596            & $5.95462*10^-$$^2$    & $5.91717*10^-$$^2$            & -$3.74466*10^-$$^4$                       & -0.628866  \\	
			14   &                         & $2.9639$            & $5.95623*10^-$$^2$    & $5.90161*10^-$$^2$           & -$5.46161*10^-$$^4$                      & -0.916958         \\
			15   &                         & 3.69431            & $5.42503*10^-$$^2$    & $5.34394*10^-$$^2$            & -$8.10862*10^-$$^4$                       & -1.49467             \\ 	\hline \hline
			
		\end{tabular}
		\caption{The percentage increase of elliptic flow of Proton due to the electromagnetic force evolution created in relativistic Pb+Pb collision with collision energy of $\sqrt{s}$ = 2.76 A TeV at centrality ranging from 20-30 \% (impact parameters in the range 6.24 fm $<$ b $<$ 9.05 fm). The change in elliptic flow $(\Delta V_2)$ is the difference between $V_2$ Theory and $V_2$ with EM.}
		\label{teblo}
	\end{table*}
	
	\subsection{The Equation of Motion}
	
	The drift velocity at each point on the freeze-out surface from the electromagnetic field evolution was calculated after solving the following force balance equation as in ~\cite{Gursoy:2014aka, Shen:2018GKM}: 
	
	\begin{equation}
	m \frac{d \vec{v}^\mathrm{\, lrf}_\mathrm{drift}}{dt} = q \vec{v}^\mathrm{,\ lrf}_\mathrm{drift} \times \vec{B}^\mathrm{\, lrf} + q \vec{E}^\mathrm{\, lrf} - \mu m \vec{v}^\mathrm{\, lrf}_\mathrm{drift} = 0
	\label{eq9}
	\end{equation}
	This gives us the local velocity due to electromagnetic effects.
	The last term in (\ref{eq9})  describes the drag force on a fluid element with mass $m$ on which some external (in this case electromagnetic) force is being exerted with $\mu$ the drag coefficient. Though the drag coefficient is 	still unclear, its value is precisely known only for  heavy quarks in $\mathcal{N} = 4$ SYM theory as stated in~\cite{Petreczky:2010, Gursoy:2014aka}.
	
	We investigated the force balance equation from which we got the drift velocity $\vec{v}^\mathrm{\,lrf}_\mathrm{drift}$ in every fluid cell along the freeze-out surface for different values of q on the equation. At the same time, we considered charges of ${\pm 1}$, ${\pm 2}$, ${\pm 1/3}$ and ${\pm 2/3}$. 
	After having the drift velocity calculated, we added it with the usual hydrodynamic flow velocity; then after, we fed it back to the framework in order to find out the particles velocity increment.

	\section{Results and Discussions}

\begin{table*}
	\centering
	\begin{tabular}{*{14}{|c}|}
		\hline
		\hline
		No. & Particle type          & Pt $(GeV)$ & $V_2$ Theory & $V_2$ with EM & $\Delta V_2$ &  \pbox{30cm}{Percentage Increase \\ of $ V_2$ (in \% ) due to \\ the electromagnetic field }  \\
		\hline
		\hline
		1   & \multirow{14.5}{*}{   Pion Plus}   & $0.00717752$           & $5.5400*10^-$$^6$   & $5.38847*10^-$$^6$         & $1.51576*10^-$$^7$                        & -2.73601         \\
		2   &                         & $0.0935073$          & $1.20967*10^-$$^3$  &  $1.18482*10^-$$^3$          & -$2.48565*10^-$$^5$                       & -2.05481            \\
		3   &                         & $0.0378994$           & $1.6268*10^-$$^4$    & $1.67079*10^-$$^4$          & $4.39942*10^-$$^6$                        & 2.70434           \\ 
		4   &                         & $0.174611$            & $4.16565*10^-$$^3$   & $4.21296*10^-$$^3$        & $4.73037*10^-$$^5$                       & 1.13557           \\
		5   &                         & $0.282125$            & $9.14852*10^-$$^3$   & $9.18976*10^-$$^3$         & $4.12446*10^-$$^5$                          & 0.450833            \\
		6   &                         & $0.417334$            & $1.58252*10^-$$^2$    & $1.5842*10^-$$^2$           & $1.68271*10^-$$^5$                      & 0.106331          \\
		7   &                         & 0.581994            & $2.37454*10^-$$^2$    & $2.3729*10^-$$^2$            & -$1.63782*10^-$$^4$                       & -0.0689745         \\
		
		8   &                         & $0.778479$          & $3.21444*10^-$$^2$  &  $3.20853*10^-$$^2$          & -$5.90906*10^-$$^5$                       & -0.183828            \\
		9   &                         & $1.01002$           & $4.02371*10^-$$^2$    & $4.01232*10^-$$^2$          & -$113938*10^-$$^4$                        & -0.283167           \\ 
		10   &                         & $1.28111$            & $4.73655*10^-$$^2$   & $4.71827*10^-$$^2$        & -$1.82811*10^-$$^4$                       & -0.385958           \\
		11   &                         & $1.59819$            & $5.30007*10^-$$^2$   & $5.27339*10^-$$^2$         & -$2.66865*10^-$$^4$                          & -0.503512            \\
		12   &                         & $1.97107$            & $5.67017*10^-$$^2$    & $5.63339*10^-$$^2$           & -$3.67824*10^-$$^4$                      & -0.6487          \\
		13   &                         & 2.41596            & $5.8028*10^-$$^2$    & $5.75364*10^-$$^2$            & -$4.91671*10^-$$^4$                       & -0.8473  \\	
		14   &                         & $2.9639$            & $5.63363*10^-$$^2$    & $5.56782*10^-$$^2$           & -$6.58154*10^-$$^4$                      & -1.16826         \\
		15   &                         & 3.69431            & $5.0082*10^-$$^2$    & $4.91452*10^-$$^2$            & -$9.36862*10^-$$^4$                       & -1.87066  \\		 	\hline \hline
			 
	\end{tabular}
	\caption{ The percentage increase of elliptic flow of Pion-Plus due to the electromagnetic force evolution created in relativistic Pb+Pb collision with collision energy of $\sqrt{s}$ = 2.76 A TeV at centrality ranging from 20-30 \% (impact parameters in the range 6.24 fm $<$ b $<$ 9.05 fm). The change in elliptic flow $(\Delta V_2)$ is the difference between $V_2$ Theory and $V_2$ with EM.}
	\label{teblo1}
\end{table*}	

\begin{table*}
	\centering
	\begin{tabular}{*{14}{|c}|}
		\hline
		\hline
		No. & Particle type          & Pt $(GeV)$ & $V_2$ Theory & $V_2$ with EM & $\Delta V_2$  &  \pbox{30cm}{Percentage Increase \\ of $ V_2$ (in \% ) due to \\ the electromagnetic field }  \\	\hline
		\hline
		1   & \multirow{14.5}{*}{Kaon Plus}   & $0.00717752$           & -$2.40631*10^-$$^6$   & -$2.42153*10^-$$^6$         & -$1.52199*10^-$$^8$                        & 0.632501         \\
		2   &                         & $0.0378994$          & -$6.55524*10^-$$^5$  &  -$6.51401*10^-$$^5$          & $4.12321*10^-$$^7$                       & -0.628995            \\
		3   &                         & $0.0935073$           & -$3.29362*10^-$$^4$    & -$3.27201*10^-$$^4$          & $2.16099*10^-$$^6$                        & -0.656113           \\ 
		4   &                         & $0.174611$            & -$5.55964*10^-$$^4$   & -$5.51048*10^-$$^4$        & $4.91587*10^-$$^6$                       & -0.884207          \\
		5   &                         & $0.282125$            & $7.68163*10^-$$^4$   & $7.73334*10^-$$^4$         & $5.17059*10^-$$^6$                          & 0.67311            \\
		6   &                         & $0.417334$            & $5.76995*10^-$$^3$    & $5.77054*10^-$$^2$           & $5.9139*10^-$$^7$                      & 0.0102495          \\
		7   &                         & 0.581994            & $1.46153*10^-$$^2$    & $1.46046*10^-$$^2$            & -$1.07354*10^-$$^5$                       & -0.0734528        \\
		
		8   &                         & $0.778479$          & $2.53318*10^-$$^2$  &  $2.52963*10^-$$^2$          & -$3.54238*10^-$$^5$                       & -0.139839            \\
		9   &                         & $1.01002$           & $3.58571*10^-$$^2$    & $3.57777*10^-$$^2$        & -$7.94344*10^-$$^5$                       & -0.22153           \\ 
		10   &                         & $1.28111$            & $4.49585*10^-$$^2$   & $4.48142*10^-$$^2$        & -$1.44262*10^-$$^4$                       & -0.320878          \\
		11   &                         & $1.59819$            & $5.20171*10^-$$^2$    & $5.17884*10^-$$^2$           & -$2.2869*10^-$$^4$                      & -0.439648            \\
		12   &                         & $1.97107$            & $5.66788*10^-$$^2$    & $5.63469*10^-$$^2$           & -$3.318679*10^-$$^4$                      & -0.585524          \\
		13   &                         & 2.41596            & $5.86183*10^-$$^2$    & $5.81607*10^-$$^2$            & -$4.57648*10^-$$^4$                       & -0.780726  \\	
		14   &                         & $2.9639$            & $5.72915*10^-$$^2$    & $5.66676*10^-$$^2$           & -$6.23967*10^-$$^4$                      & -1.08911         \\
		15   &                         & 3.69431            & $5.12312*10^-$$^2$    & $5.03334*10^-$$^2$            & -$8.97795*10^-$$^4$                       & -1.75244  \\		  	\hline  \hline
	\end{tabular}
	\caption{ The percentage increase of elliptic flow of Kaon-Plus due to the electromagnetic force evolution created in relativistic Pb+Pb collision with collision energy of $\sqrt{s}$ = 2.76 A TeV at centrality ranging from 20-30 \% (impact parameters in the range 6.24 fm $<$ b $<$ 9.05 fm). The change in elliptic flow $(\Delta V_2)$ is the difference between $V_2$ Theory and $V_2$ with EM.}
	\label{teblo2}
\end{table*}	

	This work focused on including the electromagnetic field evolution calculation on the well known iEBE-VISHNU code package for relativistic heavy-ion collisions~\cite{song:2007cvh}. We studied the percentage increase of elliptic flow of identified particles due to the electromagnetic force evolution created in relativistic Pb+Pb collision with collision energy of $\sqrt{s}$ = 2.76 A TeV using similar model set up as~\cite{Shen:2018GKM}. The results show that the electromagnetic force affects the flow harmonics with different magnitude as explained below. In heavy ion collision the magnetic field is expected to be  solenoidal fields from those flying charges. In line with this, during the beginning of the collision time, the dominant magnetic field is from the spectators, yet the participants of both incoming projectile ions contribute to the evolution of the electric and magnetic field created. In the next two sections we shall discuss on the evolution of the electromagnetic field and its effects on observables.
	
	\subsection{The Electromagnetic field evolution} 
	
	The first top of Fig.\ref{EMevolution.codes} is an illustration of the evolution of the three components of the electromagnetic field created in relativistic Pb+Pb collision with collision energy of $\sqrt{s}$ = 2.76 A TeV. The calculations were made at centrality ranging from 20-30 \% (impact parameters in the range 6.24 fm $<$ b $<$ 9.05 fm) and $ \eta_s$ ranging from -7.9 to 7.9. The three components of the electric fields in the local fluid rest frame at points on the freeze-out surface created are presented in Fig.\ref{EMevolution.codes}(a). The created electric field seen above and below of the reaction axis is quite different and this is expected due to the coulomb electric field created by the positively charged spectator particles at the beginning of the collision. This field created from the spectators is the reason for the current created in the plasma. The electric field in the z-direction is shown to be smaller than that of the two axes. This is presented \ref{EMevolution.codes}(c). The electric field evolution in the reaction plane increases as the system evolves. It is easy to notice that the transverse (x) and the longitudinal (y) components of the electric field are indistinguishable only at the very early stages of the evolution of the system. Yet $\vec E_y$ and $\vec B_y$ has been found to be in the same order through out the evolution. As it is seen on \ref{EMevolution.codes} (a), the y-component of the electric field varies steeply which dictates  the fact that at the very beginning of the evolution a large amount of net charge stays temporally in the “almond”-shaped overlapping region as explained in ~\cite{Deng_2012}. \\
	
	We found out that {\tt B$y$} is not symmetrical and dominated  {\tt B$x$} and {\tt B$z$} which can easily be seen from Fig.\ref{EMevolution.codes}(b). The sudden pop ups of {\tt B$y$} seen on Fig.\ref{EMevolution.codes}(f) dictates similar conclusion. Moreover field components {\tt B$x$}, {\tt E$x$}, and {\tt E$y$} which are as large as {\tt B$y$} are seen in different time of the system evolution. This can be due to the fluctuations of the positions of charged particles. This in-homogeneous spatial structural distribution of the electromagnetic field has also been studied in ~\cite{Deng_2012} where they have utilized  the  HIJING  model  to  investigate  the  generation  and  evolution  of  the electromagnetic fields in heavy-ion collisions.
	
\begin{figure*}
		\includegraphics[width=1\linewidth]{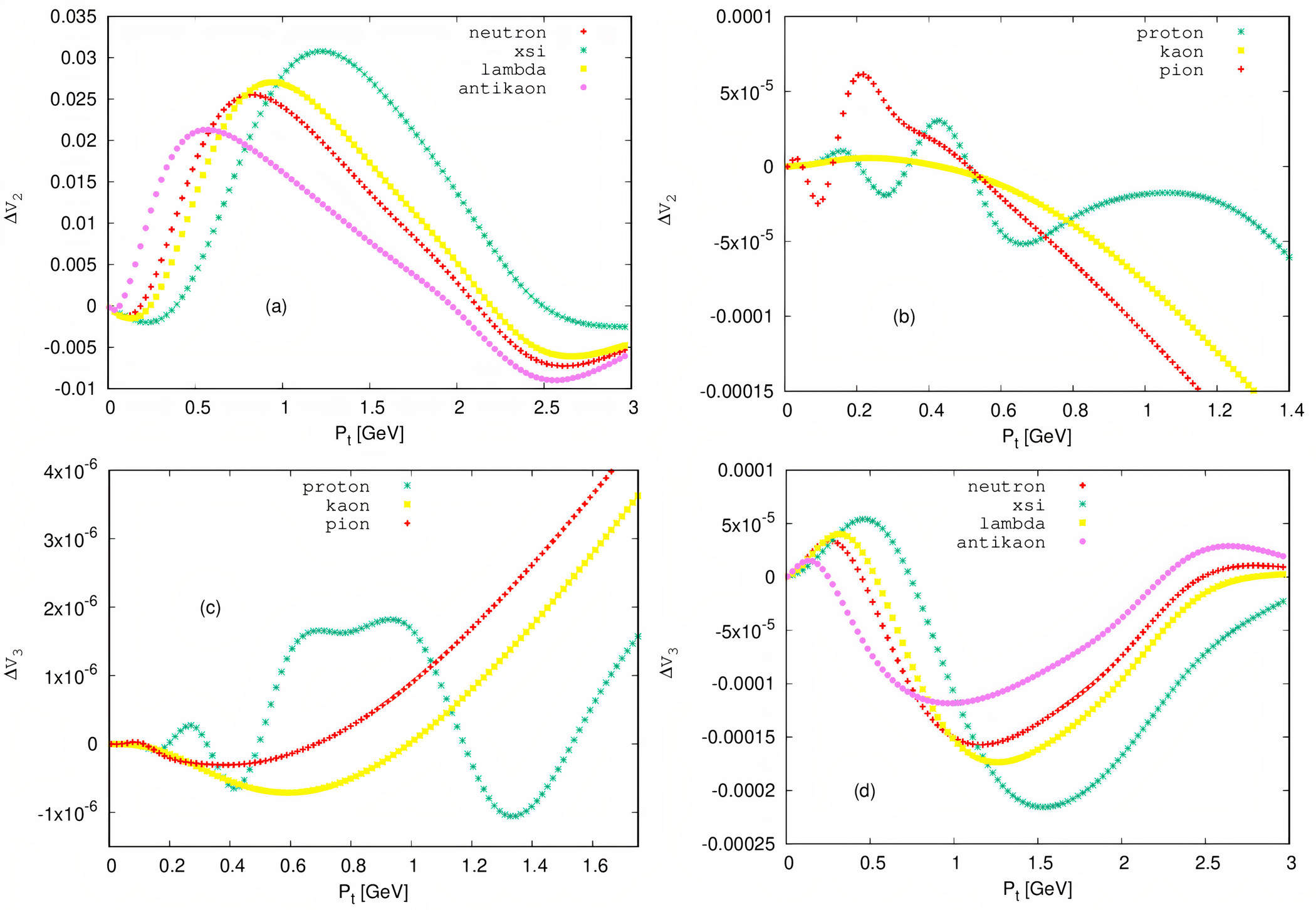}
		\caption{(Color online) Illustrate the effect of electromagnetic force on the change change in elliptic flow $\Delta V_2$ and change in triangular flow $\Delta V_3$ for some identified particles, and shows as the change in elliptic flow for heavier particles is higher than that of the lighter particles. 
		}
		\label{Best2}
		
	\end{figure*}

	By zooming out Fig.\ref{EMevolution.codes}(a) we have noticed that the electric fields perpendicular to the reaction plane shows larger gradiant than the electric fields created parallel to this plane. As explained in ~\cite{Deng_2012}, depending on how the electrical conductivity is, these outside electric fields can derive positive (negative) charged particles to move outward (toward) the reaction plane, and thus induce an electric quadruple moment which can lead to al elliptic flow imbalance between same particles of different signs; and our result have confirmed this by presenting the elliptic flow variation as given in Fig. \ref{Best2}.
		
	To provide more complete information on the early evolution of {\tt E } and {\tt B } fields, we showed in \ref{EMevolution.codes} (e) and (f) by magnifying the circled region from \ref{EMevolution.codes} (c) and (d) respectively. As expected {\tt E$_x$ } is shifted away from zero earlier which dictates the dominance of the field from the positively charged spectator particles in the beginning of the hydrodynamics. The coulomb force between the particles in the plasma creates an electric field which is against the coulomb electric field. The decreasing of the electric field in the reaction plane inturn drops the magnetic field causing Faraday electric field to evolve. Furthermore, as expected, {\tt B$z$} is smaller than {\tt B$x$} and {\tt B$y$}.
	
	As explained in the previous sections, the total electric field comes from Coulomb field of the spectators plus the plasma, the Faraday and finally the Lorentz field from the moving charges. The fluctuating electric field evolution at a fixed space time rapidity given in Fig. \ref{EMevolution.codes}(c) assures the non-uniformity of the created field. Yet, at a latter time when the spectators have already moved far away from the collision region, the contributions from the residue become important because they move much slower than the spectators. These residue can essentially slow down the decay of the transverse fields in the latter time, as seen from \ref{EMevolution.codes} and ~\cite{Deng_2012}.
	 	
	The electrical conductivity governs how fast the magnetic fields sourced initially. When it is large, the magnetic field in the plasma decays more slowly and this gives a large magnetic field for drifting particles away. In practical since the electrical conductivity is highly sensitive to temperature, a temperature dependent function should have been used. However, as given by Fig. \ref{conductivity}, our analysis used four different values of conductivity to see the order of estimate of its effect on the electromagnetic field evolution which potentially can be seen on the flow patterns. At the lower conductivity, it is easy to see that the evolution of the electric field is dominated by the $x$ component of the electric field. This is to mean that the evolved electric field contributes to the side flow rather than the longitudinal flow. This is occurred due to the coulomb electric field created by the positively charged spectator particles at the beginning of the collision. These spectators should have found enough time to evolve before the electromagnetic fields were built up by the wounded particles and the fluid it self.
	
	 As the conductivity increases, the electromagnetic field evolution in both transverse and longitudinal direction become similar. The two components of electric fields are indistinguishable at higher conductivity because the conductivity suppresses the fields.

	\subsection{The Electromagnetic field on flow harmonics}
	
	In Table \ref{teblo}, \ref{teblo1}  and \ref{teblo2} the momentum-dependent elliptic flow coefficients for proton-$P$, Pion Plus-$\pi^+$ and Kaon Plus-$K^+$ are shown. The elliptic flow for the ideal case is compared to the case with the electromagnetic field. The electromagnetic field acts on the evolution of flow that leads to the reduction of elliptic flow of these positively charged particles dominantly. The elliptic flow of protons was suppressed up to 2.37 \% of its initial value. At lower momentum they also get a percentage increase of -1.5\% from its initial value. It is different for pions, and at lower momentum the created electromagnetic field raised their elliptic flow up to 2.7\%. The effect of the electromagnetic evolution is seen to be the same for the considered three particles at higher momentum. As-far-as the system evolution is concerned, the flow harmonics values get suppressed or rouse in non uniform fashion through out the evolution. Inorder to explain the effects caused by the electromagnetic evolutions  better, every possible contributors like the drag force, electrical conductivity and gluon charge density shall be explained well. The Electromagnetic field suppressing the elliptic flow at higher momentum is also seen in ~\cite{Feng:2017FBZ} the change on the increase percentage being closer to $5\%$. \\
		
	As it is stated in Fig. \ref{Best2}, the electromagnetic field effect on $ \Delta v_2$ and $ \Delta v_3$ is small in magnitude for the positively charged and lighter particles. The change on the odd flow harmonics and the even ones is not quiet equal. The even flow harmonics get suppressed or lower in higher magnitude than the odd ones. The directed flow is influenced insignificantly. This is because the Faraday and Coulomb fields almost canceled the electric Lorentz field created in the fluid system. Nevertheless, there is still a very small contribution to the directed flow and this is from the collision axis. Meanwhile, as presented in Fig. \ref{Best2} (c) and (d), the triangular flow change is in the order of $10^-$$^4$ for neutron, xsi, antiproton, lambda and anti kaon. Yet from Fig. \ref{Best2} (a), it is easily depicted as the elliptic flow of neutron, xsi, antiproton, lambda and anti kaon is significantly influenced by the created electromagnetic evolution. This leads us to notice how the multi-directional side push of the created heavier particles is significantly influenced by the fields of electromagnetic evolution. The heavier a particle is, the higher it is pushed aside. As a result, Lambda and Xsi get larger odd flow harmonics changes than proton as found in Ref. ~\cite{Das:2017}.
	
	A hand full of other groups have studied the effect of electromagnetic field evolution on the flow of particles in relativistic heavy ion collision. They all have used different approaches making some crude assumptions to tackle the problem as we just did. All the studies ~\cite{Shen:2018GKM, Zhong:2014, Voronyuk:2011, Eric:2015} came up to similar understanding as the electromagnetic evolution has an effect on the bending of particles flow. Moreover, the result from ~\cite{Zhong:2014} on flow harmonics changes of heavier particles agrees with ours. 
	
	
	\section{Summary and Conclusions}
	
	We  have  utilized  the  iEBE-VISHNU frame work  to  investigate  the  generation  and  evolution  of  the electromagnetic fields in relativistic heavy-ion collisions. The evolution of the electromagnetic field created after a Pb+Pb relativistic collision with 20-30 centrality and collision energy of $\sqrt{s}$ = 2.76 ATeV was explored. The spatial structure of the electromagnetic field was studied and a very inhomogeneous distribution is found. We have also investigated the time evolution of the fields at the early-stage of the evolutions. We found that the residues give considerable contribution to the fields during the early-stage evolutions.  A study of the effect of this external electromagnetic field on elliptic flow of particles was investigated. Both the ideal and the electromagnetic field cases were initialized equally and the transverse profile was taken from a Monte-Carlo Glauber model initialization calculation.
		
	As a result, a maximum of $\pm$ 2.7 \% increase in elliptic flow is observed. The elliptic flow of protons was raised up to 2.37 \% from the initial value. At lower momentum, the elliptic flow of proton gets suppressed to 1.5\% of its initial value. For pions, the flow was raised by 2.7 \%. As-far-as the system evolution is concerned the flow harmonics values get suppressed or rouse in non uniform fashion through out the evolution. Inorder to explain the effects caused by the electromagnetic evolution better, every possible contributor to the change in elliptic flow should be included in functional form. 
	
	Besides, we found out that heavier particles like lambda and xsi get the higher flow harmonics increase. The directed flow is not that affected as elliptic flow because the different radial electric forces cancel out each other. In here, mass is the dominant factor than charges which is observed as particles and their anti-particles get washed aside by the field in similar fashion. The lighter particles get slight flow harmonics increase, and they had to flow off the region before getting pushed by the fields.
	
	To conclude, the present study shows that the inclusion of electromagnetic field affects the flow of particles by suppressing or rising it in non uniform fashion through out the evolution. At last, further study is needed to establish a better understanding  on the electromagnetic field evolution and its effects on the created system by softening many of the crude assumptions we made and keeping the functionality of parameters. \\ \vspace{0.5in}

	\begin{acknowledgments}
		We would like to thank Chun Shen for the briefing discussion on the codes we have used for the frame work. Moreover, we would like to forward our gratitude to Scott Pratt of Michigan State University for the valuable discussions we had on the work. Finally, we want to thank Nigus Michael for proofreading our manuscript.
	\end{acknowledgments}


\begin{thebibliography}{90}
		\bibitem{Nasim}
		M.~Nasim, S.~Shi, S.~Chatterjee, S.~Singha and V.~Roy, {Collectivity in High Energy Heavy-Ion Collisions}, Advances in High Energy Physics, vol. 2017, Article:1485353 (2017), 	
		\newblock \href {https://link.aps.org/doi/10.1155/2017/1485353} {\path{doi/10.1155/2017/1485353}},  \href{http://dx.doi.org/10.1155/2017/1485353}.
		 
			
		\bibitem{2008jna}
		The DOE/NSF Nuclear Science Advisory Committee, {The frontiers of nuclear science, a long range plan}, nucl-ex. (2008)	
	    \newblock \href {http://arXiv.org/abs/0809.3137} {\path{arXiv:0809.3137}},	\href{http://arXiv.org/abs/0809.3137}{{\tt 0809.3137}}.
	
		
		\bibitem{PefectLiquid}
		 ~RHIC, {Scientists serve up perfect liquid}, Brookhaven National Laboratory (2013), 
		 \newblock \href {https://www.bnl.gov/newsroom/news.php?a=110303} {\path{bnl.gov/newsroom/news.php?a=110303}},	\href{www.bnl.gov/newsroom/news.php?a=110303}{{\tt 0809.3137}}.
		 	
		
		\bibitem{Jacobs:2007dw}
		P.~Jacobs, {Phases of QCD: Summary of the Rutgers long range plan town meeting}, Nucl-Ex (2007),		
		\newblock \href {http://arxiv.org/abs/0705.1930} {\path{arXiv:0705.1930}}.
		
		
		\bibitem{Heinz:2013wva}
		U.~W. Heinz, {Towards the Little Bang Standard Model}, J.Phys.Conf.Ser. 455(2013) 012044.
		\newblock \href {http://arxiv.org/abs/1304.3634} {\path{arXiv:1304.3634}},
		\href {http://dx.doi.org/10.1088/1742-6596/455/1/012044}
		{\path{doi:10.1088/1742-6596/455/1/012044}}.
		
		\bibitem{Song:2008UH}
		H.~Song and U.~Heinz, {Suppression of elliptic flow in a minimally viscous quark- gluon plasma}, Phys. Lett. B 658. \newblock \href {https://arxiv.org/pdf/0709.0742} {\path{arxiv.org/pdf/0709.0742}},   \href{http://dx.doi.org/10.1016/j.physletb.2007.11.019}{\path{doi:10.1016/j.physletb.2007.11.019}}.
		
		\bibitem{Song:2011BH}
		H. ~Song, S. ~Bass, U. ~Heinz, T. ~Hirano, and C. ~Shen, {Collisions serve a nearly perfect quark-gluon liquid}, Phys.Rev.Lett. 106 (2011) 192301, \newblock \href {https://arxiv.org/pdf/1011.2783} {\path{arxiv.org/pdf/1011.2783}},
		\href {http://dx.doi.org/10.1103/PhysRevLett.106.192301}
		{\path{10.1103/physrevlett.106.192301}}.
		
		
		\bibitem{Luzum:2011}
		M. ~Luzum, {Elliptic flow at energies available at the CERN Large Hadron Collider: Comparing heavy-ion data to viscous hydrodynamic predictions},
		Phys.Rev. C83 (2011) 044911, 
		\newblock \href {https://arxiv.org/pdf/1011.5173} {\path{arxiv.org/pdf/1011.5173}},
		\href {http://dx.doi.org/10.1103/PhysRevC.83.044911}
		{\path{10.1103/PhysRevC.83.044911}}.   
		
		
		\bibitem{QGP-Lee:1974ma}
		T.~D. Lee and G.~C. Wick, {Vacuum stability and vacuum excitation in a spin-0 field theory}, Phys. Lett. B 658. \newblock \href {https://link.aps.org/doi/10.1103/PhysRevD.9.2291} {\path{doi/10.1103/PhysRevD.9.2291}},   \href{http://dx.doi.org/10.1103/PhysRevD.9.2291}.
		
		
		\bibitem{QGP-Collins:1974ky}
		J.~C. Collins and M.~J. Perry, {Superdense Matter: Neutrons or Asymptotically Free Quarks?}, PhysRevLett.34.1353. (1974) \newblock \href {https://link.aps.org/doi/10.1103/PhysRevLett.34.1353} {\path{doi/10.1103/PhysRevLett.34.1353}}, \href{http://dx.doi.org/10.1103/PhysRevLett.34.1353}. 
			
		\bibitem{Feng:2017FBZ}
		F.~Bohao, and Z.~Wang, {Effect of an Electromagnetic Field on the Spectra and Elliptic Flow of Particles},
		Phys.Rev. C95.5 (2017), 
		\newblock \href {https://arxiv.org/abs/1705.07842v1} {\path{arxiv.org/abs/1705.07842v1}},
		\href {https://link.aps.org/doi/10.1103/PhysRevC.95.054912}
		{\path{10.1103/PhysRevC.95.054912}}.   
	
		
		\bibitem{Shen:2018GKM}
		U.~Gursoy, D.~Kharzeev, E.~Marcus, K.~Rajagopal and C.~Shen, {Charge-dependent flow induced by magnetic and electric fields in heavy ion collisions}, Phys. Rev. D74 (2018) 05288,  
		\newblock \href {https://arxiv.org/abs/1806.05288} {\path{arxiv.org/abs/1806.05288}},
		\href {https://link.aps.org/doi/10.1103/PhysRevC.98.055201}
		{\path{	10.1103/PhysRevC.98.055288}}. 
		
		 
		\bibitem{Shen:2014lye}
		C.~Shen, The standard model for relativistic heavy-ion collisions and electromagnetic tomography, Doctoral dissertation, The Ohio State University, 2014.
	
	
		
	
		\bibitem {Song:2014}
		C.~Shen, Z.~Qiu, H.~Song, J.~Bernhard, S.~Bass, U.~Heinz, ''{The iEBE-VISHNU code package for relativistic heavy-ion
			collisions}'', eprint. (2014), \newblock \href {https://arxiv.org/abs/1409.8164} {\path{arxiv.org/abs/1409.8164}},
		\href{https://arxiv.org/abs/1409.8164}.
		
		
		\bibitem{Kharzeev:1996sq}
		D.~Kharzeev, "{Can gluons trace baryon number?}", Phys. Lett. B378 (1996), 
		\newblock \href {https://arxiv.org/abs/1806.05288} {\path{arxiv.org/abs/nucl-th/9602027}},
		\href {https://arxiv.org/abs/nucl-th/9602027}{{\tt arXiv:nucl-th/9602027}}.
		
		\bibitem{Gursoy:2014aka}
		U.~Gursoy, D.~Kharzeev, and K.~Rajagopal, Magnetohydro-dynamics, charged currents and directed flow in heavy ion collisions, Phys. Rev. C89(5) (2014)
		\newblock \href {http://arxiv.org/abs/1401.3805} {\path{arxiv.org/abs/1401.3805}}, 
		\href {https://link.aps.org/doi/10.1103/PhysRevC.89.054905}
      	{\path{	10.1103/PhysRevC.89.054905}}. 
		
		
		
		
		\bibitem{Kharzeev:2007jp}
		D.~E.~Kharzeev, L.~D.~McLerran, and H.~J.~Warringa, "{The Effects of topological charge change in heavy ion collisions: ‘Event by event P and CP violation’}, Nucl. Phys. A803:227-253 (2008)
		\newblock \href {http://arxiv.org/abs/0711.0950} {\path{arxiv.org/abs/0711.0950}},
		
		\href {https://link.aps.org/doi/10.1016/j.nuclphysa.2008.02.298}
		{\path{	10.1016/j.nuclphysa.2008.02.298}}. 
		
		
	
		
		\bibitem{Zhong:2014}
		Y.~Zhong, C.~Yang, X.~Cai and S.~sFeng ''{A Systematic Study of Magnetic Field in Relativistic Heavy-Ion Collisions in the RHIC and LHC Energy Regions}'', Advances in High Energy Physics,Hindawi Limited 
		\href{http://arxiv.org/abs/1408.5694v1.10.1155/2014/193039} {{\tt 10.1155/2014/193039}}.
		
		\bibitem{Voronyuk:2011}
		V.~Voronyuk, V.~Toneev, W.~Cassing, E.~Bratkovskaya, V.~Konchakovski and S.~Voloshin, {Electromagnetic field evolution in relativistic heavy-ion collisions} Phys. Rev.C.62.361 (2012)
		\newblock \href {dx.doi.org/10.1103/10.1103/physrevc.83.054911}
		 {\path{arXiv:1202.3233v1}},
		 \href{http://dx.doi.org/10.1103/physrevc.83.054911}{\path{doi:10.1103/physrevc.83.054911}}.
		 
		 
		\bibitem{Eric:2015}
		E.~Marcus, {Magnetohydrodynamics at Heavy Ion collision}, Bachelor Thesis Utrecht University
		(2015), 
		\href{http://dspace.library.uu.nl/handle/1874/317375}{\path{doi:10.1103/physrevc.83.054911}}.
		
	
		
		\bibitem{Zhong:2014}
		Y.~Zhong, C.~Yang, X.~Cai and S.~sFeng ''{A Systematic Study of Magnetic Field in Relativistic Heavy-Ion Collisions in the RHIC and LHC Energy Regions}'', Advances in High Energy Physics, Hindawi Limited 
		\href{http://arxiv.org/abs/1408.5694v1.10.1155/2014/193039} {{\tt 10.1155/2014/193039}}.
		
		\bibitem{Petreczky:2010}
		P.~Huovinen and P.~Petreczky {QCD Equation   of State and Hadron Resonance Gas} Phys. Rev. C.62.361 (2012),		
		\newblock \href {dx.doi.org/10.1016/j.nuclphysa.2010.02.015}
    	{\path{arXiv:0912.2541}},
    	\href{http://dx.doi.org/10.1016/j.nuclphysa.2010.02.015}{\path{doi:10.1016/j.nuclphysa.2010.02.015}}.
				
		
		
		\bibitem{Das:2017}
		S.~K.~Das, S.~Plumari, S.~Chatterjee, J.~Alam, F.~Scardina, and V.~Greco, {Directed Flow of Charm Quarks as a Witness of the Initial Strong Magnetic Field in Ultra-Relativistic Heavy Ion 	Collisions}, Phys. Lett. B768.260 (2017), \newblock \href {dx.doi.org/10.1016/j.physletb.2017.02.046}
		{\path{arxiv.org/abs/1608.02231}},
		\href{https://doi.org/10.1016/j.physletb.2017.02.046}{\path{doi:10.1016/j.physletb.2017.02.046}}.
		
		\bibitem{Deng_2012}
	    W.~Deng, and  X.~Huang, {Event-by-event generation of electromagnetic fields in heavy-ion collisions}, Phys.\ Rev.\ C {\bf 85}, 044907. \newblock \href {http://dx.doi.org/10.1103/PhysRevC.85.044907} {\path{arxiv.org/pdf/1201.5108}},   \href{http://dx.doi.org/10.1103/PhysRevC.85.044907}{\path{doi:10.1103/physrevc.85.044907}}.
			
		
	\end{thebibliography}
\end{document}